\documentstyle[amstex,aps,prb,epsfig]{revtex}

\begin{document}
\title{Effect of Surface Roughness on the Universal Thermal Conductance}
\author{D.~H.~Santamore and M.~C.~Cross}
\address{Condensed Matter Physics, California Institute of Technology\\
114-36, Pasadena, California 91125}
\date{\today }
\maketitle

\begin{abstract}
We explain the reduction of the thermal conductance below the predicted
universal value observed by Schwab et al. [Nature (London) 404, 974 (2000)]
in terms of the scattering of thermal phonons off surface roughness using a
scalar model for the elastic waves. Our analysis shows that the thermal
conductance depends on two roughness parameters: the roughness amplitude $%
\delta $ and the correlation length $a$. At sufficiently low temperatures
the ratio of the conductance to the universal value decreases quadratically
with temperature at a rate proportional to $\delta ^{2}a$. Values of $\delta 
$ equal to $22\%$ and $a$ equal to about $75\%$ of the width of the
conduction pathway give a good fit to the data.
\end{abstract}


\section{Introduction}

Ballistic transport in mesoscopic system has been an active area of study.
Following Landauer's approach to electronic conductance, several groups have
derived expressions for the thermal conductance due to ballistic phonon
transport in an ideal elastic beam.\cite{ACR98,RK98,B99} The formula they
derived shows that the only material and geometry dependence of the thermal
conductance arises through the long wavelength cutoff frequencies of the
elastic waves in the beam. As the temperature $T\rightarrow0$, the
conductance is dominated by the lowest few modes with zero cutoff frequency.
The conductance then takes on a {\em universal value} $K_{u}$ with the value 
$N_{0}(\pi^{2}k_{B}^{2}T)/(3h)$ with $N_{0}$ the number of modes with zero
cutoff frequency at long wavelengths, which is four for a free standing
beam. Based on these theoretical predictions, great efforts have been \ made
to observe the universal thermal conductance. Recently, Schwab et al.\cite%
{SHWR00} successfully observed the universal thermal conductance in a
suspended silicon nitride bridge. Their experiment shows a result consistent
with the universal conductance at temperatures below about $0.08$ K$.$ Above
about $1$ K the conductance rises above this value, as the modes with
nonzero cutoff frequencies become excited and contribute to the heat
transport. However in the range $0.1$ K$\sim0.4\ $K the thermal conductance
unexpectedly {\em decreases} below the universal value.

Motivated by the results of Schwab et al.\ we theoretically investigate a
likely cause of the low-temperature thermal conductance decrease. We suggest
that the conductance decrease is caused by scattering due to rough surfaces.
Recent advanced crystal growth technology guarantees very few impurities in
the material during a substrate growth, thus eliminating the possibility of
impurity scattering. On the other hand, chemical etching can produce surface
roughness on a scale of tens of nanometers, large enough to cause
significant scattering.

In this paper we use a simple scalar model for the elastic waves. We also
use a two-dimensional approximation, which is accurate at low enough
temperatures that modes with structure across the depth of the beam---the
smallest dimension in the experimental geometry---are not excited. We also
assume that the important roughness is on the sides of the beam, rather than
the top and bottom surfaces, since the horizontal surfaces are MBE grown and
have roughness at a scale of a few atomic layers, while the side faces are
chemically etched. Kambili et al.\cite{KFFL99} have used a similar model in
a numerical investigation of the effect of surface roughness on the mode
propagation.

In the following section, the details of the two-dimensional ($2$D) scalar
model are introduced, and the scattered field calculation using a Green
function approach is presented. In Sec.\ III, the scattering probabilities
and transmission coefficients are calculated and the latter is incorporated
into the modified Landauer formula for thermal conductance. In Sec.\ IV, the
thermal conductance is evaluated numerically and compared to the experiments
of Schwab et al.

\section{Mode Scattering}

\subsection{The model}

The expression for the thermal conductance $K$ of a suspended mesoscopic
beam connecting two thermal reservoirs is\cite{ACR98,RK98,B99}
\begin{equation}
K={\frac{\hbar ^{2}}{{k_{B}T^{2}}}}\sum_{m}{\frac{1}{{2\pi }}}\int_{\omega
_{m}}^{\infty }{\cal T}_{m}(\omega ){\frac{\omega ^{2}e^{\beta \hbar \omega }%
}{{(e^{\beta \hbar \omega }-1)^{2}}}}d\omega .  \label{eq:conductivityC&L}
\end{equation}%
Here the integration is over the frequency $\omega $ of the modes $m$
propagating in the beam and $\omega _{m}$ is the cutoff frequency of the $m$%
th mode. Also $\beta =1/(k_{B}T)$, $k_{B}$ is the Boltzmann constant, $T$ is
the temperature, and $\hbar $ is Planck's constant. The effect of scattering
is introduced through the transmission coefficient ${\cal T}_{m}(\omega )$:
for the ideal case with no scattering ${\cal T}_{m}=1$. Thus, the change of
the thermal conductivity due to the rough surface is obtained by finding the
transmission coefficient.

\begin{figure}[tbp]
\begin{center}
\epsfig{file=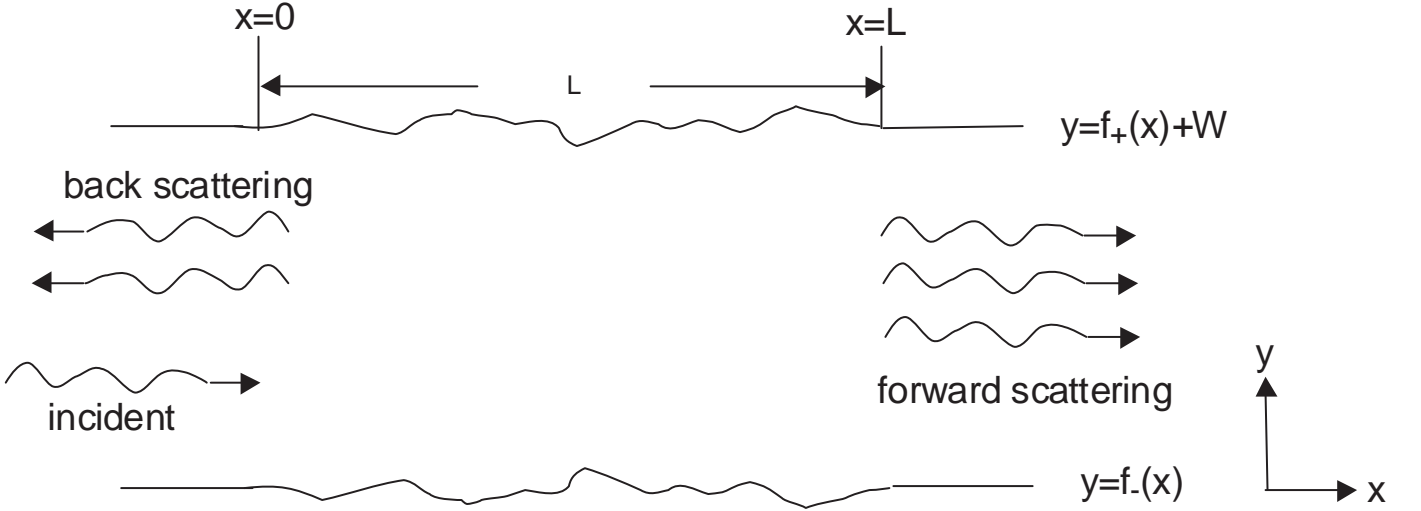}
\caption{$2$D model used for calclation of the scattering of elastic waves
by rough surfaces.}
\label{2Dmodel}
\end{center}
\end{figure}

As discussed in the introduction, we use a scalar model for the elastic
waves, and model a thin geometry at low temperatures so that a
two-dimensional calculation is adequate. Thus we consider a $2$D
wave-guide-like structure extended in the $x$ direction and bounded at $%
y=0,W $ in the absence of roughness. The waves satisfy the scalar wave
equation, and we assume Neumann boundary conditions at the edges of the wave
guide, corresponding to a stress free boundary condition for the elastic
waves. Note that Dirichlet boundaries do not support the modes with zero
cutoff frequency that are a crucial feature of the elastic problem. We
calculate the scattering process by considering an elastic wave propagating
in the wave guide in the $+x$ direction with wave vector $k_{0}$ and
entering a rough surface region of length $L$ ($0<x<L$) where the rough
boundaries are at $y=W+f_{+}(x)$ and $y=f_{-}(x)$ so that the roughness is
characterized by the functions $f_{\pm}\left( x\right) $ (see Fig.\ \ref%
{2Dmodel}). We assume that the top and the bottom roughness functions are
uncorrelated, and that $f_{\pm}\left( x\right) $ is small and is
differentiable. The incident wave $\Psi_{\text{in}}$ interacts with the
roughness, is scattered into other modes $\Psi_{\text{sc}}$, and leaves the
rough region. The total field $\Psi$\ is the sum of the incident field and
the scattered field 
\begin{equation}
\Psi=\Psi_{\text{in}}+\Psi_{\text{sc}}.  \label{eq:field}
\end{equation}
Our task is to find an expression for $\Psi_{\text{sc}}$ and hence calculate
the transmission coefficients. We do this using a Green function method.

\subsection{Green function method}

We start with the Helmholtz equation for a scalar wave at frequency $\omega$ 
\begin{equation}
\nabla^{2}\Psi\left( x,y\right) +K^{2}\Psi\left( x,y\right) =0,
\label{helmholz}
\end{equation}
where $\Psi$\ is the total field, and $K$ is $\omega/c$ with $c$ the wave
speed. Define the Green function as a solution to the point sources 
\begin{equation}
\nabla^{2}G\left( x,y;x^{\prime},y^{\prime}\right) +K^{2}G\left(
x,y;x^{\prime},y^{\prime}\right) =-\delta\left( x-x^{\prime}\right)
\delta(y-y^{\prime})  \label{Green-fcn DE}
\end{equation}
with $\left( x^{\prime},y^{\prime}\right) $\ the source coordinates and $%
\left( x,y\right) $ the observation coordinates. It is convenient to define $%
G\left( x,y;x^{\prime}y^{\prime}\right) $ such that it satisfies Neumann
boundary conditions at the {\em smoothed boundaries}, $y=0,W$ 
\begin{equation}
\partial G/\partial n|_{y=0,W}=0,  \label{b.c.}
\end{equation}
where $\hat{n}$ is the outward-pointing normal to the surface. We then
project the physical boundary conditions at the rough surfaces onto the
smoothed boundaries to calculate the scattering.

Multiplying Eq.\ (\ref{helmholz}) by $G(x,y;x^{\prime},y^{\prime})$ and Eq.\
(\ref{Green-fcn DE}) by $\Psi(x,y)$ and integrating over a volume bounded by
the position of the smoothed surfaces yields the result of Green's theorem 
\begin{align}
\Psi\left( x,y\right) & =\int_{\text{smooth}}dx^{\prime}\frac{\partial
\Psi\left( x^{\prime},y^{\prime}\right) }{\partial n^{\prime}}G\left(
x^{\prime},y^{\prime};x,y\right)  \nonumber \\
& \pm\int_{x\rightarrow\pm\infty}dy^{\prime}\left[ \frac{\partial\Psi\left(
x^{\prime},y^{\prime}\right) }{\partial x^{\prime}}G\left( x^{\prime
},y^{\prime};x,y\right) -\Psi\left( x^{\prime},y^{\prime}\right) \frac{%
\partial G\left( x^{\prime},y^{\prime};x,y\right) }{\partial x^{\prime}}%
\right]  \label{total-field}
\end{align}
where the first integral is the integration along the smoothed edges and the
second is the integration over the distant ends taken at $x\rightarrow
\pm\infty$, and we have used the boundary condition Eq.\ (\ref{b.c.}) for $G$
to eliminate a second term in the first integral.

The Green function $G\left( x,y;x^{\prime},y^{\prime}\right) $ satisfies
Eq.\ (\ref{Green-fcn DE}). Using the completeness relation, the right hand
side can be written 
\begin{equation}
-\frac{1}{2\pi}\int_{-\infty}^{\infty}e^{ik(x-x^{\prime})}dk\sum_{n}\phi
_{n}(y)\phi_{n}(y^{\prime}),
\end{equation}
where $\phi_{n}$ is the normalized transverse eigenfunction for smooth
boundaries 
\begin{equation}
\phi_{n}=N_{n}\cos\chi_{n}y
\end{equation}
with $\chi_{n}=n\pi/W,\ n=0,1,2,...$, \ and $N_{n}$ the normalization factor 
$N_{n}=\sqrt{2/W}$ for $n\neq0$ and $N_{n}=\sqrt{1/W}$ for $n=0$. The Green
function is then given by Fourier transforming 
\begin{equation}
G\left( x,y;x^{\prime},y^{\prime}\right) =\frac{1}{2\pi}\int_{-\infty
}^{\infty}dk\sum_{n}\frac{e^{ik(x-x^{\prime})}\phi_{n}(y)\phi_{n}(y^{\prime})%
}{k^{2}+(n\pi/W)^{2}-K^{2}},.  \label{Green fcn}
\end{equation}
The $k$ integral in Eq.\ (\ref{Green fcn}) is now evaluated by contour
integration. The poles corresponding to propagating waves at $k=\sqrt {%
K^{2}-(n\pi/W)^{2}}$ for $K>n\pi/W$ must be given infinitesimal imaginary
parts $\pm i\varepsilon$ to yield outgoing waves. We then have 
\begin{equation}
G\left( x,y;x^{\prime},y^{\prime}\right) =\sum_{n}\frac{ie^{ik_{n}|x-x^{%
\prime}|}\phi_{n}(y)\phi_{n}(y^{\prime})}{2k_{n}}  \label{Green Function Sum}
\end{equation}
where 
\begin{equation}
k_{n}=\left\{ 
\begin{array}{cc}
\sqrt{K^{2}-n^{2}\pi^{2}/W^{2}} & n\pi/W<K, \\ 
i\sqrt{n^{2}\pi^{2}/W^{2}-K^{2}} & n\pi/W>K.%
\end{array}
\right.
\end{equation}
Using the explicit expression for the Green function the second term in Eq.\
(\ref{total-field}) \ can be shown to be just the incoming wave $%
\Psi_{in}\left( x,y\right) $, so that 
\begin{equation}
\Psi_{sc}\left( x,y\right) =\int_{\text{smooth}}dx^{\prime}\frac {%
\partial\Psi\left( x^{\prime},y^{\prime}\right) }{\partial n^{\prime}}%
G\left( x^{\prime},y^{\prime};x,y\right)
\end{equation}

\subsection{Boundary perturbation}

In the absence of roughness the field $\Psi$ satisfies Neumann boundary
conditions at the smooth boundary, and so the scattered field would be
identically zero as expected. Correspondingly, for a rough surface with
small $f_{\pm}(x)$ we can calculate $\partial\Psi/\partial n$ at the {\em %
smoothed} surface appearing in the integral by expanding about the
stress-free rough surface.\cite{BGKP} We will present the calculation for
the rough lower surface, and simply double the scattering probabilities
assuming uncorrelated roughness on the two surfaces.

Firstly, express the unit normal vector as 
\begin{equation}
\hat{n}=-\hat{y}+f_{-}^{\prime}\left( x\right) \hat{x}.
\end{equation}
Then impose the Neumann boundary condition at $y=f_{-}(x)$ 
\begin{equation}
\text{\ }\left. \left( -\frac{\partial\Psi\left( x,y\right) }{\partial y}%
+f_{-}^{\prime}(x)\frac{\partial\Psi\left( x,y\right) }{\partial x}\right)
\right| _{y=f_{-}(x)}=0.
\end{equation}
Now expand this equation about $y=0$ in terms of $f_{-}$ and retain only
terms that are first order in $f$ and $f^{\prime}$. This gives for the
normal derivative at the smooth surface up to first order in $f,f^{\prime}$ 
\begin{equation}
\left. \partial_{n}\Psi\left( x,y\right) \right| _{y=0}=\left. \left[
f_{-}^{\prime}\partial_{x}\Psi\left( x,y\right)
-f_{-}\partial_{y}^{2}\Psi\left( x,y\right) \right] \right| _{y=0}.
\label{1st-order-bc}
\end{equation}
Thus the scattered field to first order in the roughness amplitude is 
\begin{equation}
\Psi_{\text{sc}}\left( x,y\right) \simeq\int dx^{\prime}G\left( x^{\prime
},y^{\prime};x,y\right) \left. \left[ -f_{-}\left( x^{\prime}\right)
\partial_{y^{\prime}}^{2}\Psi_{\text{in}}\left( x^{\prime},y^{\prime}\right)
+f_{-}^{\prime}\left( x^{\prime}\right) \partial_{x^{\prime}}\Psi _{\text{in}%
}\left( x^{\prime},y^{\prime}\right) \right] \right| _{y^{\prime}=0}
\label{scattered-field}
\end{equation}
where we can replace the field appearing in the integral by the incident
field $\Psi_{\text{in}}$ at this order.

It is now straightforward to insert the explicit expression for the Green
function Eq.\ (\ref{Green Function Sum}) to calculate the scattering from a
normalized incident wave entering in the $m$ th mode $\Psi_{\text{in}}\left(
x,y\right) =\Psi_{\text{m}}(x,y)=N_{m}\cos\left( \chi_{m}y\right)
e^{ik_{m}x} $: 
\begin{equation}
\Psi_{\text{sc}}\left( x,y\right) \simeq\int dx^{\prime}\sum\limits_{n}\frac{%
iN_{n}^{2}N_{m}}{2k_{n}}\cos\left( \chi_{n}y\right) e^{ik_{n}\left|
x-x^{\prime}\right| }\left[ f_{-}\left( x^{\prime}\right)
\chi_{m}^{2}+ik_{m}f_{-}^{\prime}\left( x^{\prime}\right) \right]
e^{ik_{m}x^{\prime}}.  \label{Eq_psi_sc}
\end{equation}

\subsection{Scattered field}

Outside the scattering region, we may take an asymptotic form for the
scattered field Eq. (\ref{Eq_psi_sc}) 
\begin{align}
\Psi_{\text{sc}}\left( x\rightarrow+\infty,y\right) &
=\sum_{n}e^{ik_{n}x}\cos\left( \chi_{n}y\right)
\int_{-\infty}^{\infty}dx^{\prime}\frac {iN_{n}^{2}N_{m}}{2k_{n}}\left[
\chi_{m}^{2}f_{-}\left( x^{\prime}\right) +ik_{m}f_{-}^{\prime}\left(
x^{\prime}\right) \right] e^{i(k_{m}-k_{n})x^{\prime}},  \nonumber \\
\Psi_{\text{sc}}\left( x\rightarrow-\infty,y\right) &
=\sum_{n}e^{-ik_{n}x}\cos\left( \chi_{n}y\right)
\int_{-\infty}^{\infty}dx^{\prime }i\frac{iN_{n}^{2}N_{m}}{2k_{n}}\left[
\chi_{m}^{2}f_{-}\left( x^{\prime }\right) +ik_{m}f_{-}^{\prime}\left(
x^{\prime}\right) \right] e^{i(k_{m}+k_{n})x^{\prime}}
\end{align}
giving the forward scattered field and back scattered fields respectively.
The terms in $f_{-}^{\prime}$ can be simplified by integration by parts

\begin{equation}
\int_{-\infty}^{\infty}dx^{\prime}if_{-}^{\prime}\left( x^{\prime}\right)
k_{m}e^{i\left( k_{m}\mp k_{n}\right) x^{\prime}}=k_{m}\left( k_{m}\mp
k_{n}\right) \tilde{f}_{-}\left( k_{m}\mp k_{n}\right) ,
\end{equation}
where $\tilde{f}_{-}$\ is the Fourier transform of $f_{-}$, and we have used
the fact that the roughness is confined to $0<x<L$ so that $f_{-}(\pm
\infty)=0$.

Now using $K=\omega/{c}=\sqrt{\chi_{m}^{2}+k_{m}^{2}}$, where $c$ is the
velocity of the elastic wave, we get 
\begin{equation}
\Psi_{sc}\left( x\rightarrow\pm\infty,y\right) =\sum_{n}\frac{iN_{n}N_{m}}{%
2k_{n}}\tilde{f}_{-}\left( \left( k_{m}\mp k_{n}\right) \right) \left( {K^{2}%
}\mp k_{n}k_{m}\right) \Psi_{n}\left( x,y\right) ,
\end{equation}
for the forward and backward scattered waves, expressed as a sum over
normalized waves $\Psi_{n}$.

\section{Thermal Conductance}

\label{Sec_Conductance}

Let $t_{\pm n,m}$ be the scattering amplitude from mode $m$ to $\pm n$,
where the plus sign is for forward scattering and the minus sign is for back
scattering. Then 
\begin{equation}
t_{\pm n,m}=\frac{iN_{n}N_{m}}{2k_{n}}\left( {K^{2}}\mp k_{n}k_{m}\right) 
\tilde{f}_{-}\left( k_{m}\mp k_{n}\right) .
\end{equation}
To calculate the transmission coefficient appearing in the expression for
the thermal conductance we need the energy flux scattering probabilities $%
\sigma_{\pm n,m}$ given by multiplying $|t_{n,m}|^{2}$ by the ratio of the
group velocities 
\begin{equation}
\sigma_{\pm n,m}=\frac{k_{n}}{k_{m}}\left\langle \left| t_{\pm n,m}\right|
^{2}\right\rangle ,
\end{equation}
where we can also now average over the ensemble of surface roughness
represented by the angular brackets. This finally gives 
\begin{equation}
\sigma_{\pm nm}=\frac{N_{n}^{2}N_{m}^{2}}{4k_{n}k_{m}}\left[ {K^{2}}\mp
k_{n}k_{m}\right] ^{2}\delta^{2}\tilde{g}(k_{m}\mp k_{n})L
\end{equation}
using 
\begin{equation}
\left\langle \left| \tilde{f}_{-}(k)\right| ^{2}\right\rangle =\delta ^{2}%
\tilde{g}(k)L,
\end{equation}
where $\delta^{2}\tilde{g}(k)$ is the Fourier transform of the surface
roughness correlation function with $\delta$ the roughness amplitude. For
the characterization of the rough surface, we assume a Gaussian correlation
function $g(x)=e^{-x^{2}/a^{2}}$ where $a$ is the correlation length of the
roughness, so that 
\begin{equation}
\tilde{g}(k)=\sqrt{\pi}a\exp\left[ -a^{2}k^{2}/4\right] .
\end{equation}

To calculate the thermal conductance we must recognize that not all
scattering processes decrease the heat transport. A wave entering in mode $m$
has four possible outcomes: after the scattering events it may stay in mode $%
m$ propagating forward; it may be converted to mode $n$ also propagating
forward; it may stay in mode $m$ but propagating backward; and finally it
may be converted to mode $n$ and propagating backward. The former two cases
do not change the heat transport, since each mode at frequency $\omega$
contributes the same amount to the conductance. The two back scattering
events do reduce the heat transport however. Thus the backward scattering
rate $\sigma_{-n,m}$ contributes to the reduction of the thermal conductance
while $\sigma_{+n,m}$ is the coefficient for forward scattering, and leaves
the conductance unchanged. We therefore define the conductance attenuation
coefficient per unit length of the rough surface waveguide as $%
\gamma_{m}\equiv(2/L)\sum _{n}\sigma_{-n,m}$ (where the factor of $2$ is to
include the scattering off the top surface) 
\begin{equation}
\gamma_{m}=\sum_{n}{{\frac{{(K^{2}+k_{n}k_{m})^{2}}}{{k}_{n}{k_{m}}}}\frac{%
\sqrt{\pi}N_{n}^{2}N_{m}^{2}\delta^{2}a}{2}e^{-{a^{2}}(k_{n}+k_{m})^{2}/4}.}
\label{Eq_gamma}
\end{equation}

The conductance attenuation coefficient $\gamma _{m}$ gives the exponential
decay rate of the wave in mode $m$, so that over a length $L$ the
transmission is 
\begin{equation}
{\cal T}_{m}=e^{-\gamma _{m}L}.  \label{Eq_Tm}
\end{equation}%
To calculate the thermal conductance at a given temperature, we insert Eqs.\
(\ref{Eq_gamma}),(\ref{Eq_Tm})\ into Eq.\ (\ref{eq:conductivityC&L}): 
\begin{equation}
K={\frac{\hbar ^{2}}{{k_{B}T^{2}}}}\sum_{m}{\frac{1}{{2\pi }}}\int_{\omega
_{m}}^{\infty }d\omega {\frac{\omega ^{2}e^{\beta \hbar \omega }}{{(e^{\beta
\hbar \omega }-1)^{2}}}}\exp \left[ -{\sum_{n}{\frac{{(K^{2}+k_{n}k_{m})^{2}}%
}{{k}_{n}{k_{m}}}}\frac{N_{n}^{2}N_{m}^{2}\sqrt{\pi }\delta ^{2}aL}{2}e^{-{%
a^{2}}(k_{n}+k_{m})^{2}/4}}\right]  \label{thermal-conductance}
\end{equation}

\begin{figure}[tbp]
\begin{center}
\epsfig{file=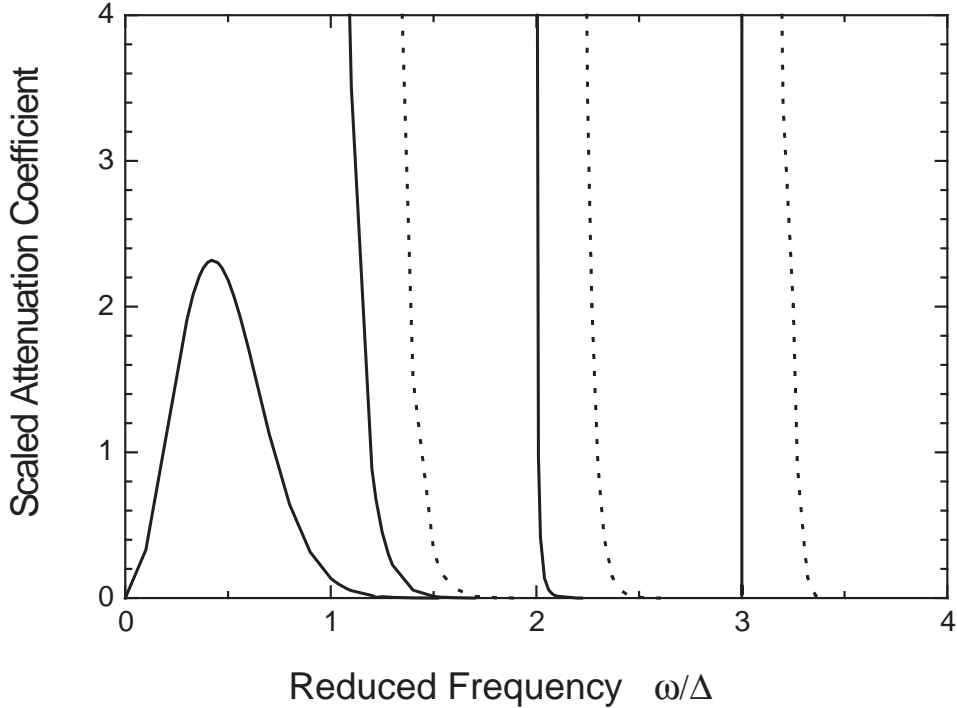}
\caption{Scaled attenuation coefficient $(W^{4}/\protect\delta^{2}aL)|%
\protect\sigma _{-n,m}|^{2}$ as a function of reduced frequency, $\protect%
\omega/\Delta$ where $\Delta=\protect\pi c/W$ with $c$ the average velocity
of the elastic waves: solid - from mode $m=0$ to mode $-n$, $n=0..3$ and
dashed - mode $m$ to mode $-m$, $m=1..3$. A value of the roughness
correlation length $a=0.75W$ was used.}
\label{attenuation-mode}
\end{center}
\end{figure}

The contributions to the conductance attenuation coefficients per unit
length $\gamma_{m}$ for the first few modes are shown as a function of the
mode frequency in Fig.\ \ref{attenuation-mode}. A transverse correlation
length of $a/W=0.75$ was used in the figure. The backscattering amplitude
from the lowest mode (mode $0$) to its reverse is 
\begin{equation}
\gamma_{00}\left( \omega,a,\delta\right) =2\pi^{1/2}\frac{a\delta^{2}}{W^{2}}%
\frac{{\omega}^{2}}{c^{2}}{e^{-a^{2}\omega^{2}/c^{2}}.}
\label{Eq_BackScatter0}
\end{equation}
This expression is finite for all frequencies. It has a maximum at a
frequency $\omega=c/a$ depending on the roughness correlation length, with a
peak value of order $(\delta^{2}/aW^{2})$. The higher modes have a diverging
back scattering proportional to $(\omega-\omega_{m})^{-1}$ at the cutoff
frequencies $\omega_{m}$. In addition each $\gamma_{m}$ has a contribution
diverging as $(\omega-\omega_{n})^{-1/2}$ at the onset of the $n$th mode.
These divergence are due to the flat spectrum at the mode cutoff
frequencies, and will also be found in a full elastic wave calculation.

At low enough temperatures only the lowest mode with $k_{0}=\omega/c$
contributes to the thermal conductance, and only the backscattering of this
mode given by Eq.\ (\ref{Eq_BackScatter0}) reduces the conductance below the
universal value. This reduction is plotted as a function of the temperature
scaled by $\hbar c/k_{B}a$ in Fig.\ \ref{mode1-scaled}. The temperature of
maximum reduction depends on the roughness correlation length $a$, whereas
both the roughness amplitude and correlation length change the magnitude of
the reduction. For small roughness, we can expand the exponential term in
Eq.\ (\ref{thermal-conductance}) and at low temperatures only small $\omega$
contributes to the integral so that the Gaussian factor may be replaced by
unity, ${\exp[}-a^{2}\omega^{2}/c^{2}]\simeq1$. This leads to 
\begin{equation}
\frac{K}{T}\simeq\frac{\pi^{2}k_{B}^{2}}{3h}\left[ 1-\frac{8\pi^{9/2}}{5}%
\frac{\delta^{2}aL}{W^{4}}\left( \frac{k_{B}T}{\hbar\Delta}\right) ^{2}%
\right] ,  \label{lowTconductance}
\end{equation}
where $\Delta=\pi c/W$ is the spacing between the mode cutoff frequencies.
Thus at low temperatures, the conductance divided by temperature should show
a quadratic temperature decrease with an amplitude depending on the
combination of roughness parameters $\delta^{2}a/W^{3}$.

\section{Comparison with experiment}

\label{Sec_Experiment}

To compare with the experiments of Schwab et al.\cite{SHWR00} we use the
following geometry and material parameters. We take a wave guide structure
of rectangular cross section with width $W=160$ nm and length $L=1$ $\mu$m.
In the experimental geometry the width varied along the length to provide
smooth junctions with the reservoirs. This was done to eliminate scattering
off abrupt changes in the geometry. We use the width at the narrowest point
as our estimate. For the length we use the length of the central portion
over which the width is fairly constant. Since the length only occurs in the
combination $\delta^{2}L$, changing the value of $L$ used will only change
the estimated value of $\delta$. We use a wave propagation speed $c=8250$
m/s which is the average of the velocity of longitudinal and transverse
elastic waves in silicon nitride.

\begin{figure}[tbp]
\begin{center}
\epsfig{file=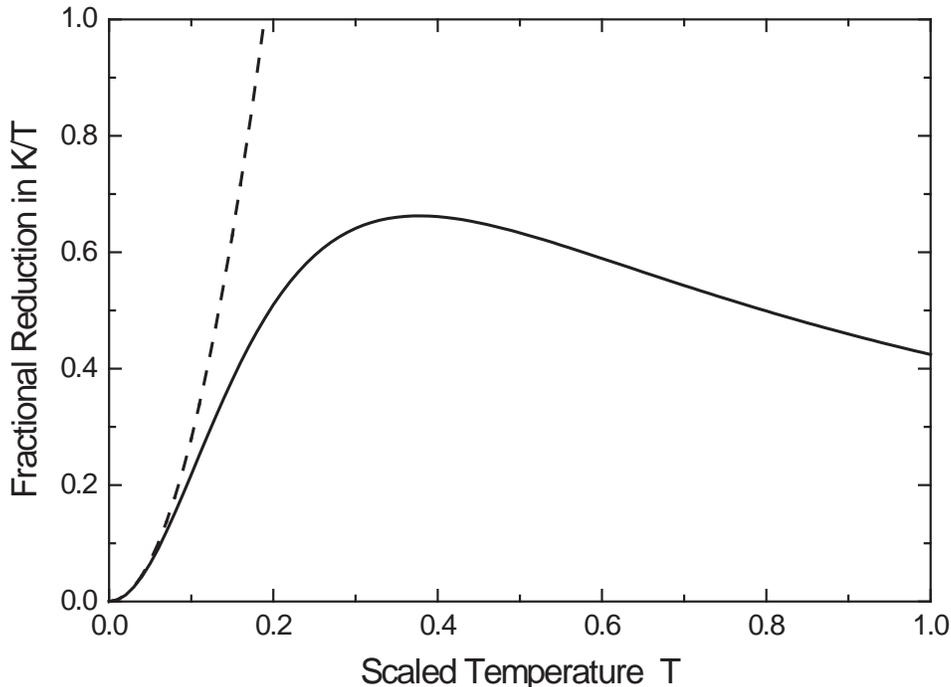}
\caption{Reduction in the thermal conductance divided by temperature due to
back scattering of the lowest mode, expressed as the ratio to the universal
conductance divided by temperature and then scaled by $aW^{2}/\protect\delta%
^{2}L$, as a function of temperature scaled by $\hbar c/k_{B}a$.}
\label{mode1-scaled}
\end{center}
\end{figure}

The roughness parameters are not known {\it a priori}. As a first attempt we
might try to estimate the combination $\delta^{2}aL/W^{4}$ from the
quadratic decrease in the thermal conductance at low temperatures, Eq.\ (\ref%
{lowTconductance}). This would give the value $a\delta^{2}L/W^{4}\sim0.05$.
However, from Fig.\ \ref{mode1-scaled} we see that the quadratic
low-temperature fit is only good up to about a quarter of the temperature of
the maximum backscattering of the first mode. If we estimate this
temperature from the minimum in the measured conductance, we find that the
data does not extend to low enough temperatures to provide a reliable fit,
and so this value can only be used as an order of magnitude. In fact our
``best fit'' (see below) over temperatures up to $1$ K corresponds to a
value $a\delta^{2}L/W^{4}$ about a factor of $4$ larger

It is interesting to use this value of the roughness parameters to estimate
the strength of the scattering of the higher modes. For example, for the
first mode, with cutoff frequency $\Delta$, and at a wave vector $\pi/W$
corresponding to a frequency $\sqrt{2}\Delta$ we find for the backscattering
into the same mode 
\begin{equation}
\gamma_{1}(k_{1}=\pi/W)L\sim16\exp\left( -\pi^{2}a^{2}/W^{2}\right) .
\end{equation}
The scattering increases for smaller wave vectors, diverging at onset as
shown in Fig.\ \ref{attenuation-mode}. Remember that the transmission
amplitude is $e^{-\gamma_{1}L}$. This means that the scattering of the
higher modes is {\em strong} over the $1\mu$ length, unless sufficiently
reduced by the exponential factor arising from the reduced roughness at
short length scales. To fit the higher temperature data using Eq.\ (\ref%
{thermal-conductance}) we will find that we need a value of $a$ comparable
to $W$. Although this strongly reduces the value of $\gamma_{1}(k_{1}=\pi/W)$%
, there remain frequency ranges where the scattering of this mode and other
modes is strong. An interesting consequence is that a significant fraction
of the thermally excited phonons at temperatures of order $1$ K are
predicted to be {\em localized} in the experiments of Schwab {\it et al}.,
with a localization length less than the length of the bridge.
Unfortunately, in this regime the estimate of the contribution to the
conductance from these modes predicted by our lowest order scattering
calculation, will not be accurate.

From Fig.\ \ref{mode1-scaled} we can suggest two mechanisms that might
account for the observed minimum in the dependence of $K/T$ on temperature.
The first mechanism ascribes the minimum in $K/T$ to the behavior of the
first mode alone, as plotted in Fig.\ \ref{mode1-scaled}. The upturn in $K/T$
arises from the reduced scattering of the lowest mode as the wave vectors of
the important modes increase with temperature. The second mechanism supposes
that the scattering of the lowest mode is responsible for the decreasing $%
K/T $ at low temperatures, but that the subsequent increase is from the
thermal excitation of the higher modes. For our ``best fit'' values of $%
a,\delta$ (see below) the results are summarized in Fig.\ \ref{individual}.
The picture is quite complicated, with both the reduced scattering of the
lowest mode and the thermal excitation of the higher modes contributing to
the rise in $K/T$ with increasing temperature. Furthermore, due to the
strong scattering of the higher modes near their cutoff frequencies, these
modes become important in the transport at a higher temperature than would
be estimated simply from their cutoff frequencies. The higher modes excited
near their threshold frequencies are localized and do not contribute
significantly to the transport.

\begin{figure}[tbp]
\begin{center}
\epsfig{file=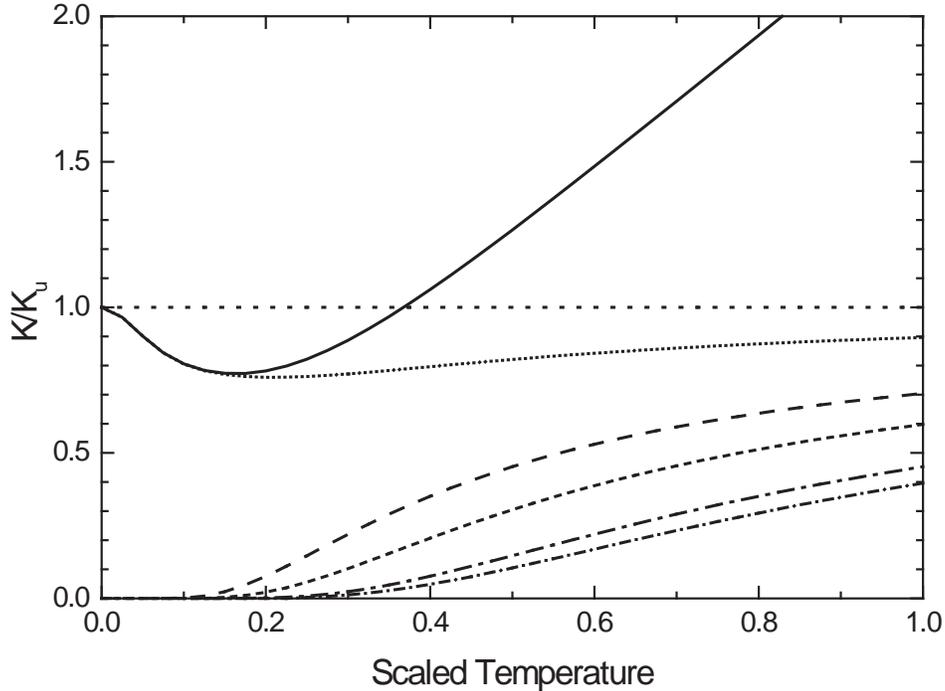}
\caption{Contributions to the thermal conductance $K$ divided by the
universal value $K_{u}$ from the first few modes for the ideal no scattering
case, and for the rough case with scattering as a function of the scaled
temperature $k_{B}T/\hbar\Delta$: solid line - total thermal conductance for
the rough surface case; dotted line and short-dotted line - conductance of
mode 0 (ideal and rough); dashed line and short-dashed line - conductance of
mode 1 (ideal and rough); dash-dotted line and short-dash-dotted line -
conductance of mode 2 (ideal and rough). Values of the roughness parameters
were $a/W=0.75$ and $\protect\delta/W=0.22$.}
\label{individual}
\end{center}
\end{figure}

\begin{figure}[tbp]
\begin{center}
\epsfig{file=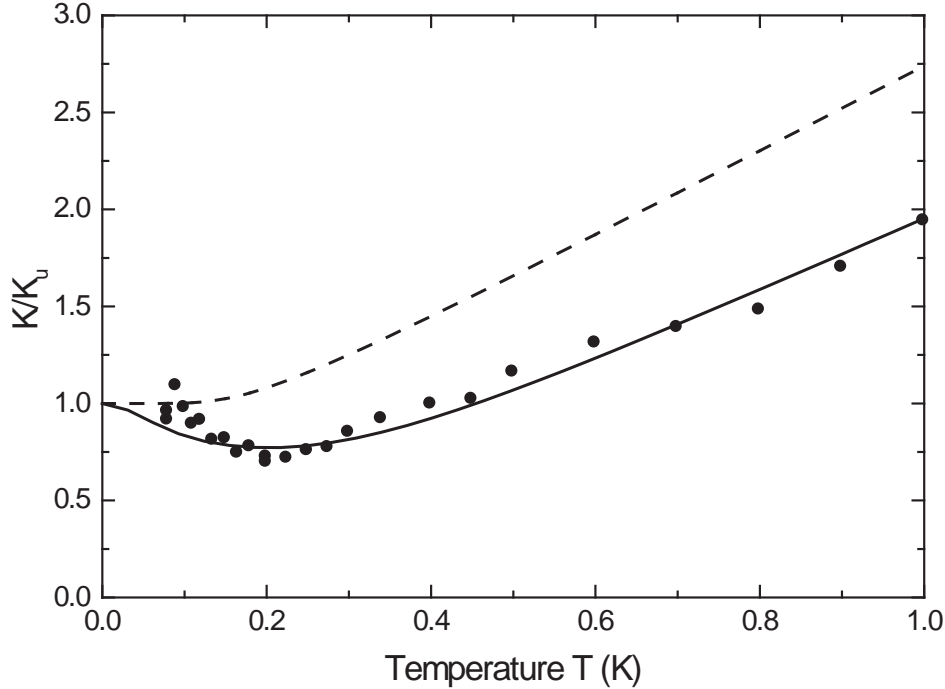}
\caption{Thermal conductance relative to the universal value $K_{u}$ as a
function of temperature for the ideal case (dotted line), the rough surface
case (solid line), and the data of Schwab et al.\ (circles). The roughness
parameters used were $a/W=0.75$, $\protect\delta/W=0.22$.}
\label{fullconductance}
\end{center}
\end{figure}

In Fig.\ \ref{fullconductance} the thermal conductance calculated using Eq. (%
\ref{thermal-conductance}) is plotted together with the ideal
(no-scattering) conductance and the measurements of Schwab et al. The
conductance is scaled such that the universal conductance appears as unity.
The roughness parameters $a/W=0.75$ and $\delta/W=0.22$ (so that $%
a\delta^{2}L/W^{4}=0.23$) were used, and yield a reasonable fit to the data.
Our $2D$ model shows the same trend as the experimental data: a decrease in
the thermal conductance below the universal value at low temperatures where
only the lowest modes are excited, then a gradually increasing conductance
as other modes are excited and the scattering of the lowest mode is reduced.
Comparison to the ideal (non-scattering) curve shows that the scattering is
important over the whole temperature range examined $(T<1$ K$)$. These
values of $\delta=35$ nm and $a=120$ nm appear reasonable when one considers
the physical process of constructing the mesoscopic bridge structure. For
example, a typical chemical etch of silicon nitride can easily produce a few
tens of nm in roughness amplitude. Electron micrographs of the actual
structure used in the experiment\cite{schwab} show roughness on scales
comparable to the ones we estimate.

There are small but systematic differences at very low temperatures, where
the conductance is dominated by the lowest modes, and the theory should be
most accurate. The discrepancy suggests that we are overestimating the
scattering at long wavelengths. A roughness spectrum $\tilde{g}(k)\sim
k^{2}e^{-a^{2}k^{2}/4}$ with a reduced amplitude at small wave numbers gives
a better fit to the data. Such a form might be physically reasonable, since
we might expect the roughness to be largest at a scale of order the minimum
dimension of the structure, and reduced at larger scales than this. However,
since the scalar model does not account for the mode structure of the
elastic beam accurately, it is probably unwise to use the discrepancies in
Fig.\ \ref{fullconductance} to make any firm deductions. Such conclusions
must await a more accurate treatment of the modes within elasticity theory.

\section{Conclusion}

We have investigated the cause of the thermal conductance decrease below the
universal value at low temperatures by employing a Green function approach
to calculate the reduced transmission of the elastic waves due to surface
roughness, and then using Landauer's formula for the thermal conductance.

At low temperatures, the conductance divided by the temperature is dominated
by the lowest mode. The scattering of this mode reduces the conductance
below the universal value with a quadratic dependence on temperature for low
temperatures with an amplitude proportional to the combination of roughness
parameters $a\delta^{2}$. As the temperature increases, higher modes begin
to play a role and the scattering of the lowest modes is reduced, so that
the conductance increases. We find that the effect of scattering is always
significant, reducing the conductance below the ideal ballistic value over
the whole temperature range we investigate $T<1$ K. Considering the
simplicity of our model our results agree well with the experiment of Schwab
et al. In future work we will present results for a full elastic theory
treatment of the thin bridge.

\begin{acknowledgments}
The authors are grateful to Keith Schwab for providing the data and an
electron micrograph of the experimental structure, and Miles Blencowe for
carefully reading the manuscript and providing useful suggestions. This work
was supported by NSF grant no. DMR-9873573.
\end{acknowledgments}

\end{document}